\documentclass{nature}

\usepackage{lineno}

\usepackage{bm}
\usepackage{url}
\usepackage{amsmath}
\usepackage{amssymb}
\usepackage{graphicx}
\usepackage{booktabs}

\title{Population-level verification of the black-hole area law with first- and second-generation black holes}

\author{Shao-Peng~Tang$^{1,\ast}$, Yin-Jie~Li$^{1,\ast}$, Yi-Zhong~Fan$^{1,2,\dagger}$.}

\begin{document}

\maketitle

\begin{affiliations}
\small
\item{Key Laboratory of Dark Matter and Space Astronomy, Purple Mountain Observatory, Chinese Academy of Sciences, Nanjing 210033, China}
\item{School of Astronomy and Space Science, University of Science and Technology of China, Hefei, Anhui 230026, China}\\
$^\ast$These authors contributed equally.\\
$^\dag$Corresponding author. Email: \texttt{yzfan@pmo.ac.cn}
\end{affiliations}

\begin{abstract}
Hawking's area theorem states that the total event-horizon area of classical black holes can never decrease.
Gravitational-wave tests of this law have so far focused on individual loud mergers, thus probing only a sparse subset of binary parameter space.
Here we perform the \emph{first} population-level test.
We exploit the decomposition of the coalescing black holes in the latest gravitational-wave catalogue into a low-spin subpopulation of stellar-collapse origin and a high-spin subpopulation assembled through hierarchical mergers of the former.
If the high-spin black holes are merger remnants, the area theorem requires their horizon areas to exceed the total pre-merger areas of the low-spin binaries.
Using parameter estimation restricted to the inspiral of 241 events, so that no merger-ringdown information enters the inference, we find that both peaks of the horizon-area distribution of second-generation black holes lie above their first-generation counterparts.
The displacement is significant at each peak ($2.1$--$3.7\sigma$), and when we tie the two subpopulations with a single common shift, an area decrease is excluded decisively ($\gtrsim5.0\sigma$).
The second law of black-hole mechanics thus holds statistically across the quasicircular, moderately spinning mergers that dominate current catalogues, which in turn underpins the robustness of our classification of stellar-collapse and hierarchical-merger black holes.
\end{abstract}

Hawking's area theorem\cite{1971PhRvL..26.1344H} states that, assuming weak cosmic censorship and the null energy condition, the total area of black hole (BH) event horizons can never decrease.
As the classical counterpart of the second law of thermodynamics, it is a cornerstone of classical general relativity (GR) and of BH thermodynamics\cite{1970PhRvL..25.1596C, 1973CMaPh..31..161B, 1973PhRvD...7.2333B, 1975CMaPh..43..199H}.
Gravitational wave (GW) observations of binary black hole (BBH) coalescences\cite{2016PhRvL.116f1102A} offer a novel way to confront the theorem with data, and the feasibility and methodology of such tests have been studied extensively\cite{2005ApJ...623..689H, 2016JCAP...10..001G, 2018PhRvD..97l4069C, 2023MNRAS.523.4113T}.
The first observational test, based on GW150914, favoured an area increase at moderate credibility\cite{2021PhRvL.127a1103I}, and independent analyses of the same event reached broadly consistent conclusions\cite{2022PhRvD.105f4042K, 2024PhRvD.110d4018C}.
More recent loud BBH events\cite{2026ApJ..1004L..22A, 2026arXiv260527225T} have yielded considerably stronger support for the theorem\cite{2025PhRvL.135k1403A, 2026SciBu..71...83T, 2026PhRvL.136d1403A}.

All of these, however, are tests of \emph{individual} events, whereas the theorem is a universal statement: a genuine proof would require verifying it for every possible coalescence, which no finite set of observations can achieve.
The practical question is therefore how far a single verified event reaches.
Strictly, it certifies the theorem at a single point in the space of merger configurations.
A modest extrapolation from that point is defensible: by the scale invariance of vacuum GR, one event plausibly stands for all total masses at fixed mass ratio and spins, although even this step is not entirely free, since new physics violating the area law would generically introduce a scale.
What a single event cannot cover is the variation \emph{within} the population it belongs to.
The BBHs detected so far are quasicircular and moderately spinning; while the coalescences in this class share qualitatively similar merger dynamics, they still span a wide range of mass ratios and spin configurations.
Letting one loud event speak for the whole class promotes that expected similarity to an assumption, when it is precisely what a test of the area law should be measuring.
Beyond this class lie regimes that current events cannot probe at all, such as near-extremal spins, large orbital eccentricity or head-on encounters, which will require dedicated tests of their own.

Between a single point and full universality, however, lies a well-posed question that can be answered now: does the area law hold across the class of quasicircular, moderately spinning mergers that dominates the current catalogues, rather than merely at the few loud events within it?
This calls for a test that is \emph{broad} rather than \emph{deep}: a population-level test, which replaces the extrapolation assumption with a direct statistical measurement over the entire class.

Such a test has recently become possible. Joint mass--spin population analyses\cite{2022ApJ...941L..39W, 2024PhRvL.133e1401L, 2026PhRvL.137b1407L, 2026SCPMA..6999562W, 2026arXiv260527226T} of the LIGO--Virgo--KAGRA (LVK) GW transient catalogues\cite{2023PhRvX..13d1039A, 2026ApJ..1004L..22A, 2026arXiv260527225T} robustly separate the coalescing BHs into two subpopulations:
a low-spin subpopulation (Pop1) whose mass function cuts off near the pair-instability gap, identified with first-generation (1g) BHs formed from stellar collapse,
and a high-spin ($\chi\sim0.7$) subpopulation (Pop2) whose spins match the natal spins of merger remnants, identified with the products of previous mergers\cite{2021NatAs...5..749G}.
Crucially, ref.~\citenum{2026arXiv260701121L} showed that the mass function of Pop2 traces, \emph{peak by peak}, the remnant-mass distribution predicted from Pop1 mergers, providing smoking-gun evidence that Pop2 objects are second-generation (2g) BHs assembled from Pop1 mergers.
Under this interpretation, the area theorem yields a sharp statistical prediction: the horizon area of each Pop2 BH must exceed the total horizon area of its (unobserved) progenitor pair, so that the peaks of the Pop2 horizon-area distribution must lie above the corresponding peaks of the Pop1 \emph{pre-merger total-area} distribution.

This construction carries a distinctive advantage: the post-merger BH is probed not through its comparatively weak ringdown, but through the loud inspiral of its \emph{subsequent} merger, where the source parameters are measured most precisely.
In this respect it inherits the spirit of the hierarchical-triple proposal of ref.~\citenum{2023MNRAS.523.4113T}, which tests the area theorem using inspiral signals alone by pairing two successive mergers along a single chain.
That proposal, however, hinges on both mergers of the same chain being individually detected and associated, which is an exceedingly rare occurrence.
Promoting the idea from individual chains to the population level removes this requirement altogether: progenitors and remnants are no longer matched event by event, but statistically, through the two subpopulations as a whole.
Here we implement such a test on the detected BBH population for the \emph{first} time.

\section*{Inspiral-only horizon areas}
We start from the 259 BBHs with false-alarm rate below $1\,{\rm yr}^{-1}$ in GWTC-5\cite{2026arXiv260527225T}, of which 222 belong to Pop1 and 37 to Pop2 according to the spin-driven classification of ref.~\citenum{2026arXiv260701121L} (Extended Data Table~\ref{tab:pop2}).
A test of the area law should not presuppose its validity.
Full inspiral--merger--ringdown (IMR) posteriors rely on waveform models that impose GR throughout the highly dynamical merger phase, which is precisely where a violation, if present, would be most likely to manifest, so these posteriors partly encode the very area increase we set out to test.
We therefore re-estimate the parameters of every event using only pre-merger data, excising the merger and ringdown with the gating-and-inpainting technique\cite{2021PhRvD.104f3030Z, 2023PhRvL.131v1402C} (Methods).
Eighteen events (11 from Pop1, 7 from Pop2) are removed because the in-band inspiral is too short for reliable inference or because the gated analysis fails to converge (Extended Data Tables~\ref{tab:pop2} and \ref{tab:rejected}); the working sample comprises 211 Pop1 and 30 Pop2 events.

For a Kerr BH with mass $m$ and dimensionless spin magnitude $\chi$, the horizon area is
\begin{equation}
    A(m,\chi)=8\pi m^2\left(1+\sqrt{1-\chi^2}\right),
    \label{eq:kerrA}
\end{equation}
where we adopt geometrized units, $G=c=1$.
For each Pop1 event we compute the pre-merger total area, $A_1+A_2=A(m_1,\chi_1)+A(m_2,\chi_2)$;
for each Pop2 event we compute the area of the primary, $A_1=A(m_1,\chi_1)$, the component most confidently associated with a previous merger remnant.
For each Pop1 event we additionally evaluate the GR-predicted remnant area (hereafter Rem1), $A_f=A(m_f,\chi_f)$, obtained by propagating its inspiral-only component parameters through the numerical-relativity-calibrated final-state fits (Methods).
Building the statistic $x\equiv\log_{10}(A/{\rm km^2})$ for each subpopulation, we find the same morphology whether the areas are taken from our inspiral-only posteriors or from the LVK full-IMR samples: each subpopulation shows two prominent over-densities, and both Pop2 structures are coherently displaced towards larger areas relative to their Pop1 counterparts (Extended Data Figure~\ref{fig:stack}).

\section*{Parametric peak inference}
The two well-separated over-densities seen in each subpopulation (Extended Data Figure~\ref{fig:stack}) suggest that a low-order Gaussian mixture in $x$ should capture this morphology, and we therefore begin with a parametric description.
We fit the single-event posteriors hierarchically (Methods), with the effective single-event prior on $x$ evaluated by a Jacobian--convolution method, and adopt a three-component mixture for Pop1 and a two-component mixture for Pop2, as favoured by the Bayesian evidences among the families we considered.
The observables are not the component means but the \emph{modes} of the mixture density, located numerically for each hyperposterior sample.

\begin{figure}
\centering
\includegraphics[width=0.96\textwidth]{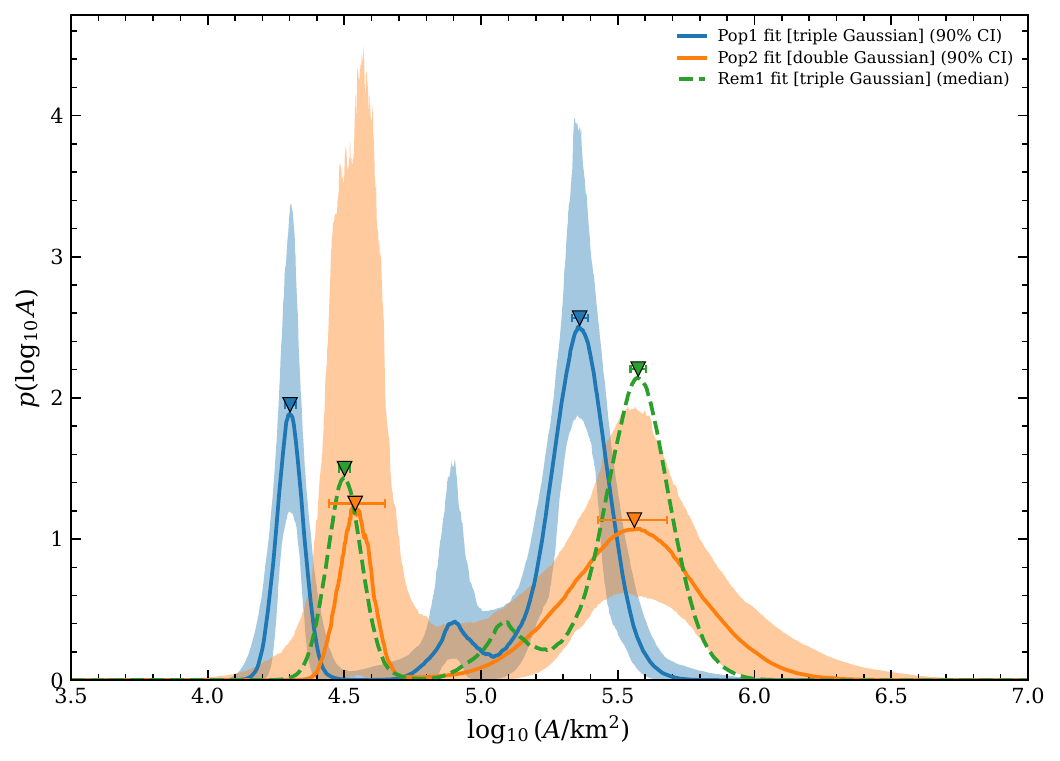}
\caption{\textbf{Parametric population distributions of the horizon-area
statistic.}
Hierarchically inferred densities of $\log_{10}A$ for Pop1 (blue; three-component Gaussian mixture of the pre-merger total horizon areas), Pop2 (orange; two-component mixture of the horizon areas of 2g BHs), and the GR-predicted Pop1 remnants Rem1 (green; three-component mixture of $A_f$).
Solid curves show the posterior median density and bands the $90\%$ credible intervals; triangles with horizontal error bars mark the posterior medians and $90\%$ credible intervals of the two peak locations.}
\label{fig:para-ppc}
\end{figure}

\begin{figure}
\centering
\includegraphics[width=0.96\textwidth]{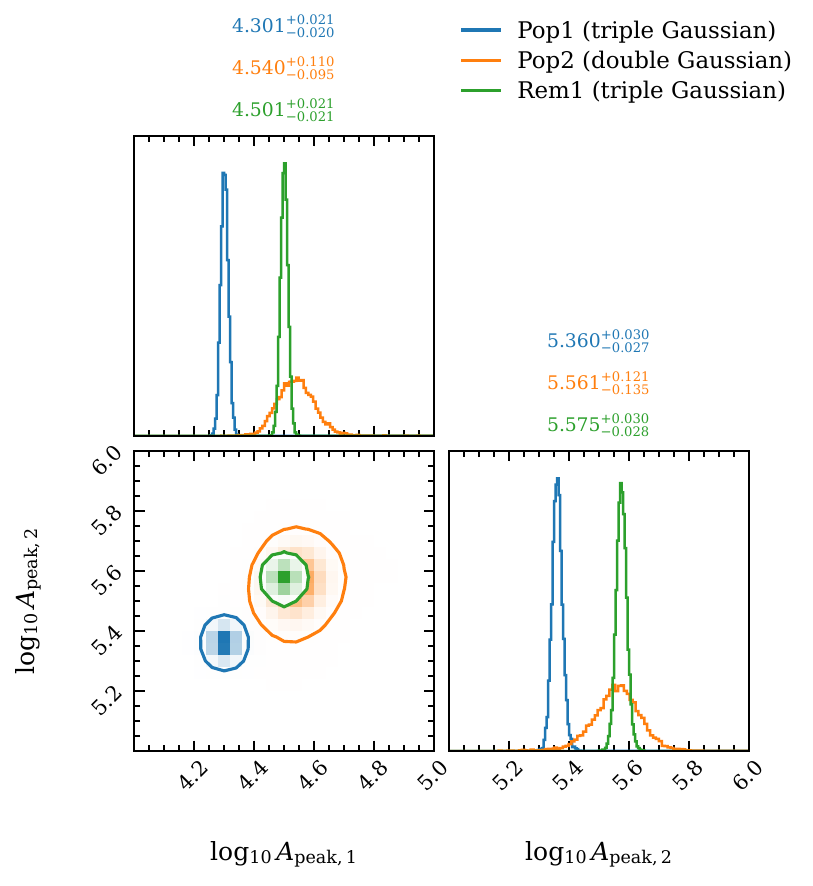}
\caption{\textbf{Joint posteriors of the peak locations from the parametric
fit.}
Joint posteriors of the two peak locations $(\log_{10}A_{\rm peak,1},\log_{10}A_{\rm peak,2})$ of the Kerr horizon-area distributions for Pop1 (blue), Pop2 (orange), and the GR-predicted Pop1 remnants Rem1 (green).
Contours enclose $90\%$ credible regions; titles above the diagonal panels give medians and $90\%$ credible intervals.
The Pop2 posterior is displaced towards larger areas in both dimensions and coincides closely with Rem1.}
\label{fig:para-peak}
\end{figure}

Figure~\ref{fig:para-ppc} shows the inferred population distributions of horizon areas.
For Pop1 we find peaks at $x_{1}^{\rm Pop1}=4.30^{+0.02}_{-0.02}$ and $x_{2}^{\rm Pop1}=5.36^{+0.03}_{-0.03}$ (all intervals are $90\%$ credible), i.e., $A=2.00\times10^{4}$ and $2.30\times10^{5}\,{\rm km^2}$.
For an equal-mass, slowly spinning pair these correspond to component masses of $\simeq9.6$ and $\simeq32\,M_\odot$, precisely the two well-known ($\sim\!10$ and $\sim\!30$--$35\,M_\odot$) peaks of the first-generation mass function\cite{2026arXiv260527226T, 2026arXiv260701121L}.
For Pop2 we find $x_{1}^{\rm Pop2}=4.54^{+0.11}_{-0.10}$ and $x_{2}^{\rm Pop2}=5.56^{+0.12}_{-0.14}$, i.e., $A=3.45\times10^{4}$ and $3.65\times10^{5}\,{\rm km^2}$, corresponding (at the expected remnant spin $\chi\simeq0.7$) to primary masses of $\simeq19$ and $\simeq62\,M_\odot$, which matches the $\sim\!17$--$19\,M_\odot$ and $\sim\!60$--$65\,M_\odot$ over-densities of the hierarchical subpopulation\cite{2026arXiv260701121L}.

\begin{figure}
\centering
\includegraphics[width=0.96\textwidth]{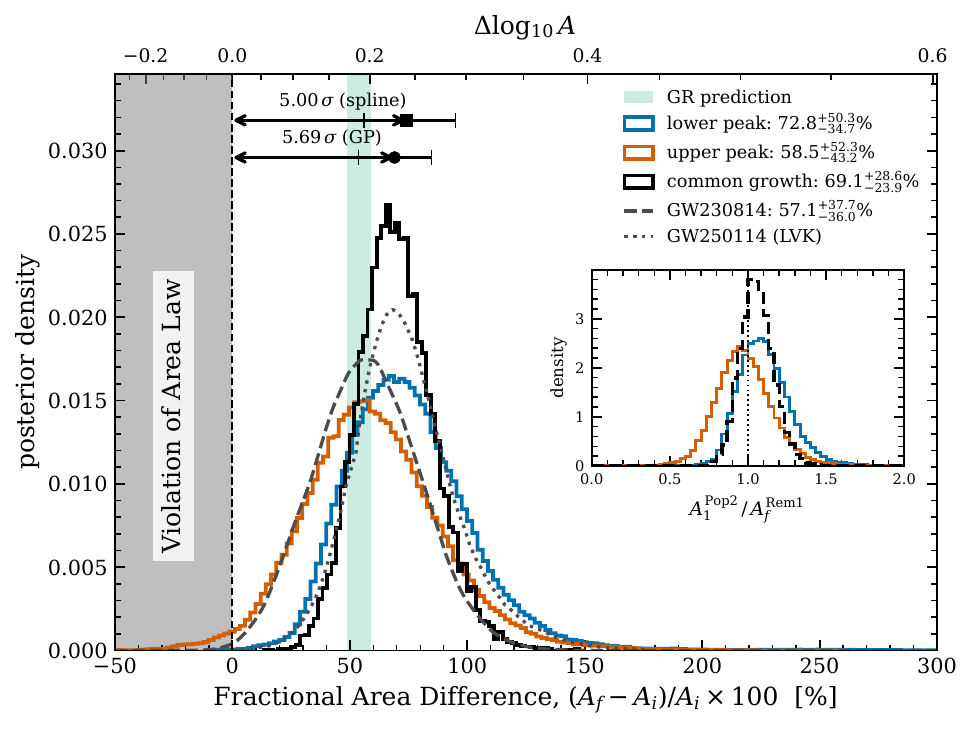}
\caption{\textbf{Fractional area differences and comparison with single-event tests.}
Posterior densities of the fractional area difference $(A_{f}-A_{i})/A_{i}$ (lower axis, in per cent; the equivalent logarithmic growth $\Delta\log_{10}A$ is marked on the upper axis) for the lower (blue) and upper (orange) peaks of the Pop2/Pop1 comparison and for the common growth factor of the joint fit (spline model, solid black).
Legends quote medians with $90\%$ credible intervals.
For the joint (common-growth) fits, the arrows at the top left (one for the spline model, one for the GP model) give the significance at which $\Delta\log_{10}A<0$ is excluded, defined as the separation between the posterior median and zero in units of the half-width of the $68.3\%$ highest-posterior-density interval.
The vertical dashed line marks zero area change, and the shaded region ($A_{f}<A_{i}$) corresponds to a violation of the area law.
For comparison, the grey dashed curve shows the fractional area change of GW230814 (derived from the orange solid line in Figure~4 of ref.~\citenum{2026SciBu..71...83T}), the grey dotted curve the LVK measurement for GW250114\cite{2025PhRvL.135k1403A}, and the green band the GR prediction of $\simeq55\%$ (Methods).
The inset shows the posterior density of the ratio $A_{1}^{\rm Pop2}/A_{f}^{\rm Rem1}$ of the observed Pop2 primary area to the GR-predicted Pop1 remnant area; it peaks near unity, showing that the Pop2 areas agree with the remnant areas expected from Pop1 mergers under GR.}
\label{fig:frac-area}
\end{figure}

Defining $\Delta_j\equiv x_{j}^{\rm Pop2}-x_{j}^{\rm Pop1}$ and drawing independently from the two hyperposteriors while preserving intra-sample peak correlations (Figure~\ref{fig:para-peak}), we obtain
\begin{align}
\Delta_1 &= +0.24^{+0.11}_{-0.10}\ {\rm dex},
\quad P(\Delta_1>0)=0.9995\ (3.3\sigma),\nonumber\\
\Delta_2 &= +0.20^{+0.12}_{-0.14}\ {\rm dex},
\quad P(\Delta_2>0)=0.9839\ (2.1\sigma).\nonumber
\end{align}
These admit two complementary joint statements.
The conservative one is the \emph{conjunction} probability that the area law holds simultaneously at both peaks, $P(\Delta_1>0\,\wedge\,\Delta_2>0)=0.9832$ ($2.1\sigma$).
The second quantifies the \emph{combined evidence} of the two peaks, treated as quasi-independent measurements, against the null hypothesis of no area increase:
combining the one-sided $p$-values $p_1=5\times10^{-4}$ and $p_2=1.6\times10^{-2}$ via Fisher's method gives $\chi^2=-2(\ln p_1+\ln p_2)=23.5$ with four degrees of freedom, i.e., $p=1.0\times10^{-4}$ ($3.7\sigma$).
The peak Kerr areas grow by $+72.8^{+50.3}_{-34.7}\%$ and $+58.5^{+52.3}_{-43.2}\%$, with the $90\%$ credible intervals of both peaks excluding zero.
Figure~\ref{fig:frac-area} confronts these population-level shifts with the two most stringent single-event measurements of the fractional area change to date.
The GW250114 posterior\cite{2025PhRvL.135k1403A} is centred at essentially the same value as our lower peak, while the median of the GW230814 measurement (derived from the orange solid line in Figure~4 of ref.~\citenum{2026SciBu..71...83T}) coincides with the central value of our upper peak.
The population-level test uses no merger-ringdown information and yet lands precisely where the deepest single-event ringdown tests do, which constitutes a nontrivial cross-validation between two independent probes of the same physics.

\section*{Model-independent confirmation}
The Gaussian mixture of the previous section is chosen by inspection of the stacked $x\equiv\log_{10}A$ posteriors (Extended Data Figure~\ref{fig:stack}), so the recovered peak locations and their displacement could in principle be artefacts of that assumed functional form.
We therefore repeat the entire hierarchical analysis with two non-parametric models of the log-density, $p(x)$, that assume no particular shape for the distribution: a natural cubic spline model of the log-density, in which $\ln p(x)\propto f(x)$ with $f$ a cubic spline through $N$ evenly spaced nodes whose values carry independent weakly informative Gaussian priors (the \emph{spline} model); and a Gaussian-process model of the log-density, in which the node values are drawn from a zero-mean squared-exponential Gaussian process (the \emph{GP} model).
Both are normalized over the fitting range and positive by construction, and neither prescribes the number of peaks (Methods).

With the spline model ($N=20$ nodes for Pop1 and $N=10$ for Pop2) the Pop1 peaks lie at $4.310^{+0.016}_{-0.018}$ and $5.352^{+0.029}_{-0.019}$, and the Pop2 peaks at $4.478^{+0.076}_{-0.055}$ and $5.504^{+0.126}_{-0.069}$; with the GP model ($N=60$ nodes for both) the corresponding values are $4.300^{+0.024}_{-0.022}$ and $5.362^{+0.044}_{-0.049}$ for Pop1 and $4.545^{+0.110}_{-0.104}$ and $5.530^{+0.108}_{-0.110}$ for Pop2 (Extended Data Figure~\ref{fig:nonp-peak}).
Both schemes reproduce the morphology of the parametric fit (Figure~\ref{fig:para-ppc}) and its peak locations (see Extended Data Figure~\ref{fig:nonp-ppc}), and both recover a positive shift in both peaks: equivalent Gaussian significances of $3.7\sigma$ and $3.4\sigma$ (lower/upper) with the spline model and $2.8\sigma$ and $2.1\sigma$ with the GP model (Extended Data Table~\ref{tab:sigma}).
The peak locations are stable as the node count is varied ($N=10$--$20$ for the spline, $40$--$60$ for the GP).
The increase in horizon area is therefore a property of the population as a whole and not an artefact of the parametrization or of a few individual events.

The same figure (Extended Data Figure~\ref{fig:nonp-ppc}) carries a second, independent piece of information.
Reconstructing the Rem1 distribution with the same non-parametric models yields peaks at $4.501^{+0.013}_{-0.015}$ and $5.570^{+0.048}_{-0.043}$ (spline) and $4.503^{+0.025}_{-0.024}$ and $5.571^{+0.048}_{-0.045}$ (GP), in close agreement with the Pop2 peaks reported above.
The Pop2 primary-area distribution thus tracks this predicted \emph{remnant} distribution peak by peak.
Two consequences follow.
First, the coherent displacement of both Pop2 peaks is what GR requires, in magnitude and not merely in sign, which is difficult to produce with a systematic error: a systematic in the mass or spin recovery would have to reproduce the GR remnant map at both peaks simultaneously.
Second, the two peaks are displaced by nearly the same amount, suggesting that a single scale governs the growth across the entire population.

\section*{A common area growth}
That suggestion has a clear physical basis.
For a quasicircular merger of slowly spinning BHs with mass ratio $q$, radiated mass fraction $\epsilon\simeq0.05$ and remnant spin $\chi_{\rm f}\simeq0.7$\cite{2005PhRvL..95l1101P, 2009PhRvD..79b4003S}, the horizon-area ratio is
\begin{equation}
\mathcal{R}\equiv\frac{A(M_{\rm f},\chi_{\rm f})}{A(m_1,0)+A(m_2,0)}
=(1-\epsilon)^{2}\,\frac{(1+q)^{2}}{1+q^{2}}\,
\frac{1+\sqrt{1-\chi_{\rm f}^{2}}}{2},
\label{eq:ratio}
\end{equation}
where the Pop1 progenitor spins ($\chi\lesssim0.2$) modify the denominator by less than $1\%$.
For near-equal-mass pairings ($q\simeq1$), equation~(\ref{eq:ratio}) gives $\mathcal{R}\simeq1.55$.
More importantly, over the mass-ratio and spin ranges actually occupied by Pop1, $\mathcal{R}$ varies very little.
We quantify this in two complementary ways (Methods).
First, modelling the distribution of the per-event predicted growth of the Pop1 events, $r_{A}\equiv A_f/(A_1+A_2)-1$, with a single Gaussian gives $r_{A}=0.553^{+0.019}_{-0.019}$.
Second, a forward prediction that does not use the individual events, drawing mass ratios from the underlying Pop1 mass-ratio distribution\cite{2026arXiv260701121L} at low spins ($\chi_{1,2}\le0.2$), returns a $90\%$ range $r_{A}\in[0.487,\,0.591]$ with a median of $0.555$.
GR therefore predicts that every first-generation merger grows its horizon area by close to $55\%$, with a population-wide full scatter of only $\sim\!0.1$ in $r_{A}$, i.e., $\sim\!0.03$~dex in $\log_{10}A$: several times smaller than the $\sim\!0.1$~dex statistical uncertainty of the peak-wise displacements.

A single common growth is therefore a well-motivated model, and the peak-by-peak agreement between the Pop2 areas and the Pop1 remnant areas in Figure~\ref{fig:para-ppc} and Extended Data Figure~\ref{fig:nonp-ppc} supports it empirically.
We accordingly fit both subpopulations \emph{jointly}: they share a single non-parametric density, described by the spline model ($N=20$ nodes) or the GP model ($N=60$ nodes), and the Pop2 samples are displaced by one shared parameter, $x^{\rm Pop2}\to x^{\rm Pop2}-\Delta\log_{10}A$ (Methods).
The joint fit gives a population-wide horizon-area growth of $\Delta\log_{10}A=0.228^{+0.041}_{-0.039}$~dex and hence $r_{A}=69.1^{+28.6}_{-23.9}\%$ with the GP model, and $\Delta\log_{10}A=0.240^{+0.049}_{-0.048}$~dex with the spline model (Figure~\ref{fig:frac-area}).
Pooling all 241 events substantially tightens the constraint relative to the peak-wise measurements, and $\Delta\log_{10}A>0$ is established at $5.7\sigma$ with the GP model and at $5.0\sigma$ with the spline model.

\section*{Consistency with general relativity}
The measured shifts are not merely positive; they agree quantitatively with the GR mapping between progenitors and remnants.
Both GR benchmarks derived in the previous section lie inside the $90\%$ credible intervals of the peak-wise increases, $72.8^{+50.3}_{-34.7}\%$ and $58.5^{+52.3}_{-43.2}\%$, and of the common growth, $69.1^{+28.6}_{-23.9}\%$ (Figure~\ref{fig:frac-area}).
The same agreement is visible directly in the horizon-area domain: the ratio of the Pop2 areas to the GR-predicted Pop1 remnant areas Rem1 is consistent with unity (inset of Figure~\ref{fig:frac-area}), and the Rem1 peaks coincide with the Pop2 peaks in both the parametric and the non-parametric reconstructions (Figures~\ref{fig:para-ppc} and \ref{fig:para-peak}; Extended Data Figures~\ref{fig:nonp-ppc} and \ref{fig:nonp-peak}).

The central values of the lower-peak ($72.8\%$) and common-growth ($69.1\%$) measurements lie somewhat above the GR expectation of $\simeq55\%$, whereas the upper peak ($58.5\%$) sits essentially on it.
None of these offsets is significant, since the GR value lies within $\sim\!1\sigma$ of every posterior, and their direction is readily understood.
Component spin magnitudes are generally less well constrained than mass parameters in GW parameter estimation, and truncating the data before merger reduces the available signal-to-noise ratio, further degrading the spin measurement.
Moreover, GW observations primarily constrain the effective aligned-spin parameter $\chi_{\rm eff}$, leaving a degeneracy between the component spin magnitudes and their tilt angles.
Under the default parameter-estimation prior, which is uniform in spin magnitude, the spins of the intrinsically low-spin Pop1 binaries ($\chi\lesssim0.2$; ref.~\citenum{2026arXiv260701121L}) tend to be shifted upward towards the prior median, thereby biasing their inferred horizon areas downward.
Conversely, for Pop2 primaries with characteristic spins of $\chi\simeq0.7$, the same prior tends to shift the inferred spins slightly downward and hence the horizon areas upward.
Both effects act in the same direction and push the measured displacements slightly above the GR expectation.
A population-informed spin prior would be expected to reduce this offset, but we deliberately retain the agnostic parameter-estimation prior so that the test does not import the spin information on which the Pop1/Pop2 classification itself rests.
The data are thus consistent with the second law of BH mechanics operating across the population, at the quantitative level predicted by GR.
This quantitative success, in turn, demonstrates the robustness of the classification of the coalescing BHs into stellar-collapse and hierarchical-merger subpopulations\cite{2026arXiv260701121L}, on which the present test is built.

\section*{Discussion}
The interpretation of this result rests on several assumptions and potential caveats, which we now examine in turn.
First, the inference is performed on the detected sample without forward-modelling selection effects.
This is a minor limitation because the key observable is the \emph{relative} displacement between two subpopulations subject to nearly identical selection.
To manufacture a spurious ${\sim}0.2$~dex coherent shift of \emph{both} peaks, selection would have to act systematically more strongly on Pop2 than on Pop1, a differential bias for which we can identify no plausible mechanism.
Second, the Pop1/Pop2 assignment is probabilistic rather than deterministic.
Our observables, however, are the \emph{modes} of the population density, which are less sensitive to a small admixture of misclassified events: a modest contaminating fraction mainly redistributes weight into the tails and broadens the components, leaving the dominant peak locations nearly unchanged.
Moreover, to the extent that misclassification does shift the peaks, its direction is fixed by the sample-size asymmetry (211 versus 30): at any given misclassification rate, contaminants flow predominantly from the large Pop1 sample into the small Pop2 sample, dragging the Pop2 peaks towards Pop1 and thereby \emph{diluting} rather than enhancing the measured shift.
The same holds for 1g$+$2g pairings in which the observed primary is not the merger remnant.
Our results are therefore conservative in this respect.
Third, the progenitors of the Pop2 primaries are not the observed Pop1 binaries themselves, so no event-by-event association is possible; the comparison is necessarily distributional.
Two considerations make this legitimate.
The subpopulation decomposition of ref.~\citenum{2026arXiv260701121L} is driven by the \emph{spin} distributions, with no reference to horizon areas; that the Pop2 mass function subsequently reproduces, peak by peak, the remnant-mass distribution predicted from Pop1 mergers emerged as an \emph{a posteriori} corroboration rather than an input to the classification, so the area ordering tested here is not built in by construction.
Finally, the observed 2g BHs may have grown by accretion between the two mergers, which would inflate their areas beyond what the merger alone supplies.
Two independent observations limit this.
Significant accretion would necessarily alter the spin away from the remnant value, yet the Pop2 spin distribution remains centred on the expected $\chi\simeq0.7$\cite{2026arXiv260701121L}.
Any substantial accreted mass would also blur and displace the peak-by-peak correspondence between the Pop2 mass function and the Pop1 remnant-mass distribution, whereas that correspondence is observed to hold.
Significant accretion is therefore disfavoured on both counts.

Our result admits a dual reading.
Assuming GR, the observed area ordering --- including its magnitude --- provides independent, area-domain confirmation of the hierarchical origin of the high-spin subpopulation; assuming the hierarchical origin, it verifies Hawking's area law at the population level.
The statistical error budget of the peak-wise analysis is dominated by the small Pop2 sample ($N=30$): its peak uncertainties are $0.10$--$0.13$~dex, a factor of $\sim\!4$--$6$ worse than the $0.02$--$0.03$~dex achieved for Pop1.
Pooling all $241$ events into the joint fit overcomes this limitation: motivated by the nearly constant area growth that GR predicts across the Pop1 population, we can tie the two subpopulations by a single common shift, measured to be $\sim\!70\%$, which excludes an area decrease at $\gtrsim 5\sigma$.
With the O5 run and next-generation detectors, the 2g BH sample will grow severalfold, driving the independent peak-wise test into the $5\sigma$ regime and subjecting the joint fit to a far more stringent test of its agreement with general relativity, while enabling searches for anomalous sub-classes, for example highly precessing or eccentric mergers, in which new physics near merger could manifest as an anomalous area deficit.
Population-level and golden-event tests are complementary: the former bounds violations occurring in any sizable fraction of ordinary mergers, while the latter probes single events in depth.
Together they are turning the second law of black-hole mechanics from a theorem of classical relativity into a precision-tested property of the astrophysical black-hole population.

\section*{Methods}
\subsection{Event sample and subpopulation assignment}
The parent sample is the set of 259 BBHs in GWTC-5 with false-alarm rate below $1\,{\rm yr}^{-1}$\cite{2026arXiv260527225T}.
Subpopulation membership is taken from ref.~\citenum{2026arXiv260701121L}, which assigns 37 events to the high-spin subpopulation Pop2 (Extended Data Table~\ref{tab:pop2}) and the remaining 222 to the low-spin subpopulation Pop1.
Quality control removes 18 events in total: 7 from Pop2, marked by daggers in Extended Data Table~\ref{tab:pop2}, and 11 from Pop1, listed in Extended Data Table~\ref{tab:rejected}.
The working sample therefore consists of 30 Pop2 and 211 Pop1 events (241 in total).
Throughout, the Pop1 statistic is the \emph{pre-merger total} area of an observed 1g$+$1g binary and the Pop2 statistic is the area of a single 2g BH.
In addition, from each Pop1 event we derive a third statistic, Rem1, the GR-predicted remnant area $A_f=A(m_f,\chi_f)$ of that 1g$+$1g binary, computed from its inspiral-only parameters through the numerical-relativity-calibrated final-state fits.
Rem1 is used only as the GR benchmark against which the observed Pop2 areas are compared, and never enters the tests as an input.

\subsection{Inspiral-only parameter estimation}
Following the gating-and-inpainting technique\cite{2021PhRvD.104f3030Z, 2023PhRvL.131v1402C} as implemented in \textsc{PyCBC}\cite{2019PASP..131b4503B}, we excise a $1$-s data segment beginning at $t_{\rm gate}=t_{\rm ref}-10\,t_M$, where $t_{\rm ref}$ is the trigger time reported by the LVK search and $t_M\equiv GM/c^3$ is set by the median detector-frame total mass $M$ of the released posterior.
We use the \textsc{IMRPhenomXPHM} waveform model\cite{2021PhRvD.103j4056P, 2025PhRvD.111j4019C} and adopt the analysis settings and priors of the corresponding LVK analyses, with two exceptions: the coalescence time is assigned a Gaussian prior (width $10$~ms)\cite{2025PhRvL.135k1403A}, and the sky position of each event is fixed to the maximum-likelihood value from the LVK release posterior.
Calibration uncertainties are neglected.
Noise power spectral densities are taken from the LVK data releases, and strain data are obtained from the GW Open Science Center.
Gated likelihoods are evaluated with \textsc{PyCBC}, and posterior sampling is performed with \textsc{Bilby}\cite{2019ApJS..241...27A} and \textsc{Dynesty}\cite{2020MNRAS.493.3132S} using 2000 live points.
Source-frame masses are obtained from the luminosity distances assuming the Planck15 cosmology\cite{2016A&A...594A..13P}, and horizon areas are evaluated sample-by-sample from equation~(\ref{eq:kerrA}).

\subsection{Hierarchical Bayesian inference}
Denoting the population probability density of $x=\log_{10}(A/{\rm km^2})$ as $p(x\,|\,\bm{\Lambda})$, the likelihood of the catalogue is
\begin{equation}
\mathcal{L}(\{d\}\,|\,\bm{\Lambda})\propto\prod_{i=1}^{N}\frac{1}{n}\sum_{k=1}^{n}
\frac{p\!\left(x_{i}^{(k)}\,|\,\bm{\Lambda}\right)}{\pi\!\left(x_{i}^{(k)}\right)},
\label{eq:hier}
\end{equation}
where $n=5000$ posterior samples are used per event and $\pi(x)$ is the effective single-event prior, i.e., the probability density induced on $x$ by the parameter-estimation priors of each event.
To guard against Monte Carlo noise in the per-event likelihood estimates, we monitor the variance of each estimate through its effective sample size\cite{2022arXiv220400461E, 2023MNRAS.526.3495T} and penalize hyperparameter samples whose total variance exceeds a preset threshold.
For the joint fit, which uses all events, the threshold on the summed variance is $1$; in the independent fits this budget is split between the two subpopulations in proportion to their sizes, giving $7/8$ for Pop1 and $1/8$ for Pop2.
In practice the total variance stays well below these thresholds throughout the posterior, so the results are insensitive to this criterion.

\subsection{Effective prior on the horizon-area statistic}
Writing the Kerr area of equation~(\ref{eq:kerrA}) as $A_i=k\,m_i^2\,g(\chi_i)$, with $k=8\pi(GM_\odot/c^2)^2$ and $g(\chi)=1+\sqrt{1-\chi^2}$, and expressing the source-frame masses through the detector-frame chirp mass $\mathcal{M}$, mass ratio $q\le1$ and luminosity distance $d_L$, the two statistics decompose additively in the log domain,
\begin{align}
\ln A_1 &= \ln k + u + v + w + s_1,\\
\ln(A_1+A_2) &= \ln k + u + w + h,
\end{align}
where $u=2\ln\mathcal{M}$, $v=\tfrac{2}{5}\ln(1+q)-\tfrac{6}{5}\ln q$, $w=-2\ln[1+z(d_L)]$, $s_i=\ln g(\chi_i)$, and $h=v+\ln R$ with $R=g(\chi_1)+q^{2}g(\chi_2)$.
Since the parameter-estimation priors on $(\mathcal{M},q,\chi_1,\chi_2,d_L)$ are mutually independent, the density of the Pop2 statistic is the convolution of the densities of $u$, $v$, $w$, and $s_1$, and that of the Pop1 statistic is the convolution of the densities of $u$, $w$, and $h$.

Each of $u$, $v$, $w$, and $s$ is a strictly monotone function of a single parameter whose prior cumulative distribution function (CDF) is available in closed form (or by numerical quadrature for distance priors uniform in comoving volume or in the source frame).
The density of each term is therefore represented on a fine uniform grid in $\ln A$, with the probability mass in each bin given by the CDF difference between the mapped bin edges, i.e., the Jacobian density integrated over the bin.
This treatment is accurate up to the binning resolution and, in particular, automatically handles the integrable square-root singularity of the spin-factor density as $\chi\to0$.
For the compound term $h$, we discretize $q$ into bins equally spaced in $v$, weighted by the prior CDF of $q$; conditional on $q$, $R$ is the sum of the two independent variables $g(\chi_1)$ and $q^2 g(\chi_2)$, whose binned distributions are convolved on a common linear grid, the resulting conditional distribution of $h=v+\ln R$ is mapped onto the $\ln A$ grid, and the marginal distribution of $h$ follows as the $q$-weighted sum.
The component distributions are combined by FFT convolution and normalized, and the prior density follows as $\pi(x)=\ln10\times p(\ln A)$, evaluated at each posterior sample by linear interpolation.
The prior specifications are reconstructed from the stored configuration of each event's analysis; for the standard settings these are $p(\mathcal{M})\propto\mathcal{M}$ and $p(q)\propto(1+q)^{2/5}q^{-6/5}$ (uniform in the detector-frame component masses), uniform spin magnitudes, and the released distance prior.

\subsection{Parametric mixture models}
The stacked distributions (Extended Data Figure~\ref{fig:stack}) display clear substructure in each subpopulation, so we adopt a low-order Gaussian mixture in $x\equiv\log_{10}A$ as a flexible parametric description.
We model Pop1 with a three-component and Pop2 with a two-component Gaussian mixture in $x$.
These choices are supported by a comparison of the Bayesian evidences among the mixture families we considered (double/triple Gaussian, double skew-normal, double Student-$t$): 
for Pop1 the three-component Gaussian is favoured ($\Delta\ln\mathcal{Z}\geq3.1$ with respect to every two-component alternative), whereas all candidate models for Pop2 have comparable evidences, so we adopt the simpler two-component Gaussian.
The means are ordered ($\mu_1<\mu_2<\mu_3$) to break label switching.
We take flat priors on the means, $\mu_j\in[3.7,6.5]$~dex for Pop1 and $\mu_j\in[4.1,6.5]$~dex for Pop2, and log-uniform priors on the widths, $\sigma_j\in[0.02,4.2]$~dex for Pop1 and $\sigma_j\in[0.02,3.6]$~dex for Pop2.
For each hyperposterior sample we locate the local maxima of $p(x\,|\,\bm{\Lambda})$ numerically on a fine grid, identifying ${\rm peak}_1<{\rm peak}_2$ with the two dominant modes of both subpopulations.
For Pop1 we report the outermost (dominant) modes, which carry the majority of the probability mass, while the middle component accounts for a minor feature between them.
The full hyperposteriors are given in Extended Data Figures.~\ref{fig:pop1-corner} and \ref{fig:pop2-corner} (and, for the Rem1 benchmark, in Extended Data Figure~\ref{fig:rem1-corner}):
for Pop1, $\mu_1=4.302^{+0.021}_{-0.020}$, $\mu_2=4.92^{+0.33}_{-0.07}$, $\mu_3=5.361^{+0.031}_{-0.028}$~dex with $\sigma_1=0.049^{+0.024}_{-0.022}$, $\sigma_2=0.080^{+0.243}_{-0.057}$, $\sigma_3=0.100^{+0.035}_{-0.057}$~dex;
for Pop2, $\mu_1=4.54^{+0.13}_{-0.10}$ and $\mu_2=5.56^{+0.12}_{-0.14}$~dex with $\sigma_1=0.046^{+0.166}_{-0.025}$ and $\sigma_2=0.244^{+0.164}_{-0.099}$~dex.

\subsection{Non-parametric models}
As a cross-check that is free of any assumed peak structure, we model the log-density of $x\equiv\log_{10}A$ with two independent non-parametric schemes, each normalized over the fitting range $x\in[x_{\min},x_{\max}]$.
Both schemes are used in the independent and in the joint analyses described below.
The domain $[x_{\min},x_{\max}]$ is set by the range of the (subsampled) single-event posterior samples in $x$, padded by $0.3\,$dex on each side and rounded outward to the nearest $0.1\,$dex; the density is taken to be zero outside this domain.
In the joint analysis the domain is further extended so that the shifted coordinate $x-\Delta \log_{10}A$ also lies within it for the full prior range of $\Delta \log_{10}A$.

The first approach uses a cubic spline.
We describe the log-density as a natural cubic spline through $N$ nodes $x_i$ evenly spaced over $[x_{\min},x_{\max}]$,
\begin{equation}
  p(x\,|\,\{s_i\}) =
  \frac{\exp\!\big[\,f(x;\{s_i\})\,\big]}
       {\displaystyle\int_{x_{\min}}^{x_{\max}}
        \!\exp\!\big[\,f(x';\{s_i\})\,\big]\,{\rm d}x'},
  \label{eq:spline}
\end{equation}
where $f$ is the spline interpolant with $f(x_i)=s_i$, so that the density is positive by construction and normalized over the fitting range.
The overall amplitude is degenerate with the normalization, so we fix the two end nodes, $s_1=s_N=0$, and sample the remaining $N-2$ node values with truncated Gaussian priors, $s_i\sim\mathcal{N}(0,\sigma_s^2)$ on $[-30,30]$.
The prior width is scaled with the node spacing $\Delta x=(x_{\max}-x_{\min})/(N-1)$ as $\sigma_s=\sigma_{\rm ref}\sqrt{\Delta x/\Delta x_{\rm ref}}$ with $\sigma_{\rm ref}=3$ at $N=10$, chosen empirically so that the prior predictive density retains a comparable degree of flexibility as the number of nodes is varied ($N=10,15,20$).

As an alternative, we use a Gaussian-process model.
We place $N=60~(40)$ nodes $x_j$ evenly over $[x_{\min},x_{\max}]$ and model the log-density as a natural cubic spline through the node values $\bm{f}$,
\begin{equation}
  p(x\,|\,\bm{f}) =
  \frac{\exp\!\big[\,\mathcal{S}(x;\bm{f})\,\big]}
       {\displaystyle\int_{x_{\min}}^{x_{\max}}
        \exp\!\big[\,\mathcal{S}(x';\bm{f})\,\big]\,{\rm d}x'},
  \label{eq:gp}
\end{equation}
with $\mathcal{S}(x_j)=f_j$.
The node values are given a zero-mean Gaussian-process prior with an overall amplitude $\sigma$ and a squared-exponential correlation kernel, $K_{jk}=\exp[-(x_j-x_k)^2/2\ell^2]$, so that $\sigma$ sets how far the density may depart from flat and $\ell$ sets the smallest resolvable feature in $x$.
We sample in whitened (non-centred) coordinates,
\begin{equation}
  \bm{f} = \sigma\,L(\ell)\,\bm{u},\qquad
  L(\ell)L(\ell)^{\!\top}=K(\ell),\qquad
  u_j\sim\mathcal{N}(0,1),
  \label{eq:whiten}
\end{equation}
with $L$ the Cholesky factor, which removes the strong $\bm{f}$--$(\sigma,\ell)$ degeneracy and keeps the sampled prior unit-normal.
We truncate $u_j$ at $|u_j|\le6$ and adopt log-uniform hyperpriors, $\sigma\sim\mathrm{LogU}(0.1,20)$ and $\ell\sim\mathrm{LogU}(\ell_{\min},3.0)$ with $\ell_{\min}=\max(0.05,\,0.8\,\Delta x)$, where $\Delta x$ is the node spacing; the lower bound on $\ell$ prevents the process from asking for structure finer than the node grid can represent.
Unlike the spline model, whose independent node priors leave the overall additive level of the log-density unconstrained, the zero-mean Gaussian-process prior already pins this level, so all $N$ node values are sampled.

\subsection{GR-predicted fractional area increase}
Before introducing the joint analyses (next subsection), we motivate a single common shift with three complementary observations.

First, in the independent fits the reconstructed population distribution of the Pop1 remnants, Rem1 ($A_f$), is closely consistent with that of Pop2 ($A_1$): their density estimates and peak locations overlap within the posterior uncertainties (Figure~\ref{fig:para-ppc} and Extended Data Figure~\ref{fig:nonp-ppc}).
This is the qualitative expectation if Pop2 black holes are the merger products of Pop1 binaries.

Second, we quantify the intrinsic spread of the area growth that GR predicts across the Pop1 population.
Drawing mass ratios from the reconstructed underlying Pop1 mass-ratio distribution (Figure~1 of ref.~\citenum{2026arXiv260701121L}) and assuming low spins ($\chi_{1,2}\le0.2$), we propagate each binary through the numerical-relativity-calibrated final-state fits to obtain the remnant mass $m_f$ and spin $\chi_f$, and hence the predicted fractional area growth $r_{A}=A_{f,\exp}/(A_1+A_2)-1$.
The resulting distribution is narrow, with a $90\%$ credible range $r_{A}\in[0.487,\,0.591]$ and a median of $0.555$.
Since this spread ($\sim0.1$ in $r_A$, i.e., $\sim0.03$~dex in $\log_{10}A$) is several times smaller than the statistical uncertainty of the peak-wise displacements, treating the growth as a single common shift is a well-motivated model.

Third, we confirm this expectation directly on the observed events.
For every Pop1 event we evaluate $m_f$ and $\chi_f$ from the same final-state fits, using the inspiral-only component parameters as inputs, and compute the per-event predicted growth $r_{A}$.
Fitting the resulting single-event posteriors hierarchically with a single Gaussian population model yields a tight constraint, $r_{A}=0.553^{+0.019}_{-0.019}$, whose central value agrees with the forward prediction above.
To verify that this narrow constraint is not an artefact of the assumed single-Gaussian shape, we repeat the fit with two more flexible descriptions of the $r_A$ distribution: an asymmetric Gaussian and a $20$-node spline model.
Both recover essentially the same location and width, with $90\%$ credible ranges $r_A\in[0.532,0.574]$ and $r_A\in[0.542,0.564]$ (spline).
The inferred population is thus consistent, independently of the assumed functional form, with a single common growth.

\subsection{Independent and joint analyses}
We reconstruct the horizon-area distributions in two complementary ways.
In the \emph{independent} analysis, Pop1 and Pop2 are fitted with separate densities whose peak locations are compared a posteriori; this makes no assumption about how the two subpopulations are related and yields the peak-wise displacements $\Delta_j$.
In the \emph{joint} analysis, motivated by the near-constant GR-predicted growth derived above, both subpopulations are described by the \emph{same} non-parametric density, with the Pop2 samples displaced by a single shared parameter; this pools all events to constrain one growth factor and thereby sharpens the measurement.
We carry out the joint analysis with both non-parametric schemes, i.e., the spline model and the Gaussian-process model.

Concretely, the shared parameter is a rigid translation of the Pop2 coordinate,
\begin{equation}
  x^{\rm Pop2} \longrightarrow x^{\rm Pop2} - \Delta \log_{10}A,
  \qquad \Delta \log_{10}A\sim {\rm Uniform}(-1.0,\,1.0),
  \label{eq:shift}
\end{equation}
so that Pop1 events enter unshifted while each Pop2 area is shifted back by $\Delta \log_{10}A$ onto the Pop1 (progenitor) coordinate before being evaluated under the shared density.
Because the map is a constant translation in $x$, its Jacobian is unity and no reweighting of the likelihood is needed.
If Pop2 BHs are the remnants of Pop1 mergers, $\Delta \log_{10}A$ is the population-level logarithmic area growth per merger; the area law requires $\Delta \log_{10}A>0$.

\section*{Data availability}
The strain data and noise power spectral densities used in this work are publicly available from the Gravitational Wave Open Science Center.
The inspiral-only posterior samples and the area fitting posteriors generated in this study will be available on Zenodo.

\section*{Code availability}
The analysis uses the publicly available packages \textsc{PyCBC}\cite{2019PASP..131b4503B}, \textsc{Bilby}\cite{2019ApJS..241...27A}, \textsc{Dynesty}\cite{2020MNRAS.493.3132S}.
The scripts implementing the population fitting and figure generation will be released alongside the data.

\section*{Acknowledgements}
This work is supported by the National Natural Science Foundation of China under Grants Nos. 12303056, 12588101, 12233011, and 12503059.
This research has made use of data or software obtained from the Gravitational Wave Open Science Center, a service of LIGO Laboratory, the LIGO Scientific Collaboration, the Virgo Collaboration, and KAGRA. LIGO Laboratory and Advanced LIGO are funded by the United States National Science Foundation (NSF) as well as the Science and Technology Facilities Council (STFC) of the United Kingdom, the Max-Planck-Society (MPS), and the State of Niedersachsen/Germany for support of the construction of Advanced LIGO and construction and operation of the GEO600 detector. Additional support for Advanced LIGO was provided by the Australian Research Council. Virgo is funded, through the European Gravitational Observatory (EGO), by the French Centre National de Recherche Scientifique (CNRS), the Italian Istituto Nazionale di Fisica Nucleare (INFN) and the Dutch Nikhef, with contributions by institutions from Belgium, Germany, Greece, Hungary, Ireland, Japan, Monaco, Poland, Portugal, Spain. KAGRA is supported by Ministry of Education, Culture, Sports, Science and Technology (MEXT), Japan Society for the Promotion of Science (JSPS) in Japan; National Research Foundation (NRF) and Ministry of Science and ICT (MSIT) in Korea; Academia Sinica (AS) and National Science and Technology Council (NSTC) in Taiwan of China.

\section*{Author contributions}
Y.-Z.F. supervised the project and provided guidance. S.-P.T. and Y.-J.L. conceived the idea and performed the data analyses. S.-P.T. wrote the initial draft. All authors discussed the results and contributed to revising the manuscript.

\section*{Competing interests}
The authors declare no competing interests.

\renewcommand{\refname}{References}
\bibliographystyle{sn-nature}
\bibliography{refs.bib}

\clearpage
\setcounter{figure}{0}
\setcounter{table}{0}
\renewcommand{\thefigure}{\arabic{figure}}
\renewcommand{\thetable}{\arabic{table}}
\renewcommand{\figurename}{Extended Data Figure}
\renewcommand{\tablename}{Extended Data Table}
\section*{Extended Data}

\begin{table}[htbp]
\centering
\caption{\textbf{The 37 Pop2 events of ref.~\citenum{2026arXiv260701121L}.}
Daggers ($\dagger$) mark the seven events removed by the quality control.}
\label{tab:pop2}
\begin{tabular}{lll}
\\ \toprule
GW170729            & GW230820\_212515          & GW240519\_012815 \\
GW190517\_055101    & GW230914\_111401          & GW240527\_183429$^{\dagger}$ \\
GW190519\_153544    & GW230922\_040658$^{\dagger}$ & GW240615\_160735 \\
GW190521            & GW230928\_215827          & GW240622\_004008 \\
GW190602\_175927    & GW231001\_140220$^{\dagger}$ & GW240824\_205609$^{\dagger}$ \\
GW190620\_030421    & GW231005\_021030$^{\dagger}$ & GW240920\_073424 \\
GW190706\_222641    & GW231028\_153006          & GW241011\_233834 \\
GW191109\_010717    & GW231029\_111508          & GW241110\_124123 \\
GW230601\_224134    & GW231110\_040320          & GW241113\_163507 \\
GW230704\_212616$^{\dagger}$ & GW231118\_005626 & GW241127\_061008 \\
GW230706\_104333    & GW231123\_135430$^{\dagger}$ & GW241225\_082815 \\
GW230723\_101834    & GW240515\_005301          & GW241230\_233618 \\
GW230814\_061920    &                           &                  \\
\bottomrule
\end{tabular}
\end{table}

\begin{table}[htbp]
\centering
\caption{\textbf{The 11 Pop1 events removed by the quality control.}}
\label{tab:rejected}
\begin{tabular}{lll}
\\ \toprule
GW190521\_074359 & GW230819\_171910 & GW231230\_170116 \\
GW191105\_143521 & GW231026\_130704 & GW241109\_115924 \\
GW230627\_015337 & GW231113\_150041 & GW250116\_015318 \\
GW230709\_122727 & GW231127\_165300 &                  \\
\bottomrule
\end{tabular}
\end{table}

\begin{table}[htbp]
\centering
\caption{\textbf{Equivalent Gaussian significance $\sigma$ of $P(\Delta_{j}>0)$ for the peak shifts (lower peak / upper peak).}}
\label{tab:sigma}
\begin{tabular}{lccccc}
\\ \toprule
Pop2 $\backslash$ Pop1  & \shortstack{spline (10)} & \shortstack{spline (15)} & \shortstack{spline (20)} & \shortstack{GP (40)} & \shortstack{GP (60)} \\
\hline
spline (10)     & 3.70 / 3.39 & 3.51 / 3.29 & 3.69 / 3.40 & -- & -- \\
spline (15)     & 3.21 / 2.33 & 3.13 / 2.15 & 3.02 / 2.32 & -- & -- \\
spline (20)     & 3.27 / 2.49 & 3.22 / 2.34 & 3.19 / 2.43 & -- & -- \\
GP (40)         & --          & --          & --          & 2.76 / 2.18 & 2.81 / 2.17 \\
GP (60)         & --          & --          & --          & 2.77 / 2.08 & 2.82 / 2.07 \\
\bottomrule
\end{tabular}
\end{table}

\begin{figure}
\centering
\includegraphics[width=0.96\textwidth]{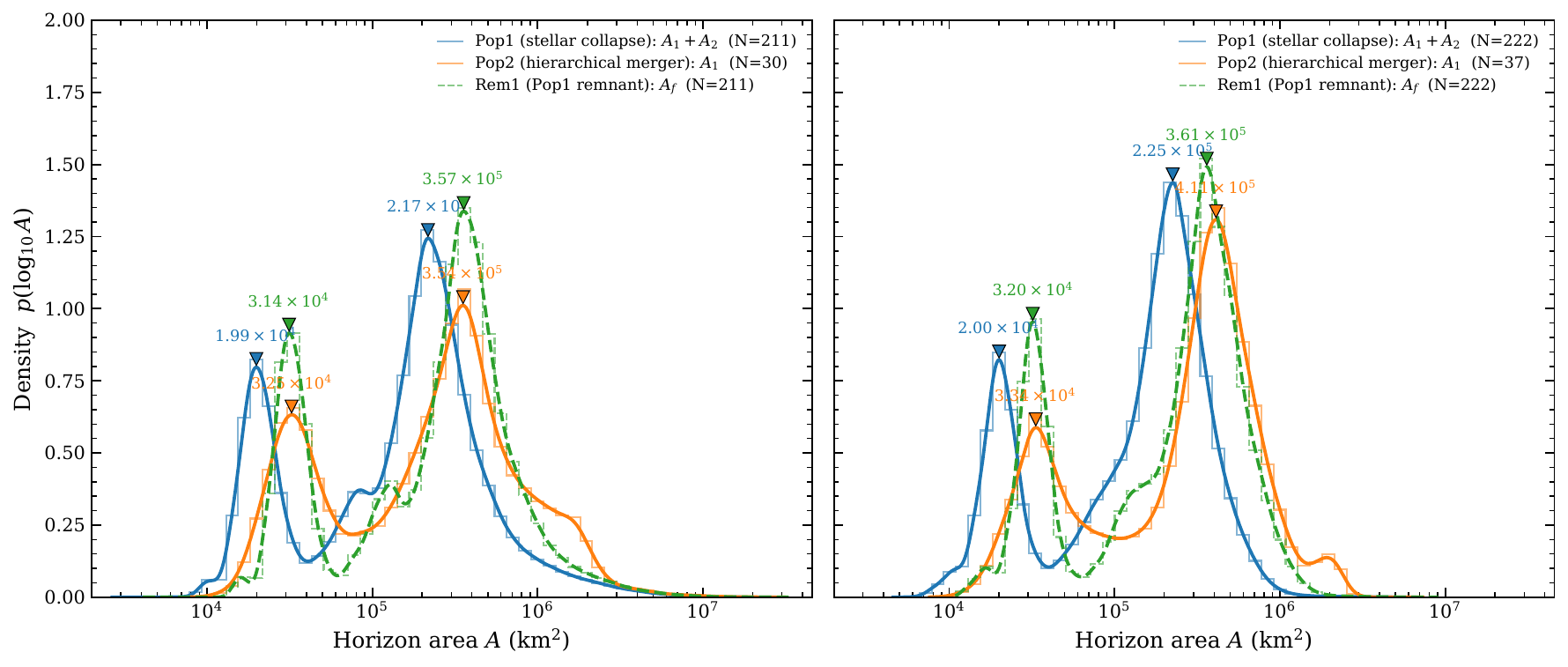}
\caption{\textbf{Stacked posterior distributions of the Kerr horizon-area statistic for the two subpopulations.}
Blue curves show Pop1, associated with a stellar-collapse origin, for which the statistic is the total pre-merger horizon area, $A(m_1,\chi_1)+A(m_2,\chi_2)$. 
Orange curves show Pop2, associated with a hierarchical-merger origin, for which the statistic is the primary horizon area, $A_1=A(m_1,\chi_1)$.
Green curves show the GR-predicted Pop1 remnants (Rem1), for which the statistic is the remnant horizon area $A_f=A(m_f,\chi_f)$.
\emph{Left}: Inspiral-only results from this work, based on 211 Pop1 and 30 Pop2 events; \emph{right}: corresponding results obtained from the LVK full-IMR posterior samples, based on 222 Pop1 and 37 Pop2 events.
Solid curves denote kernel density estimates and triangles mark their local maxima.
In both panels the two Pop2 over-densities are coherently shifted towards larger horizon areas relative to their Pop1 counterparts.
This morphology motivates the low-order Gaussian mixture description adopted in the main text.}
\label{fig:stack}
\end{figure}

\begin{figure}
\centering
\includegraphics[width=0.48\textwidth]{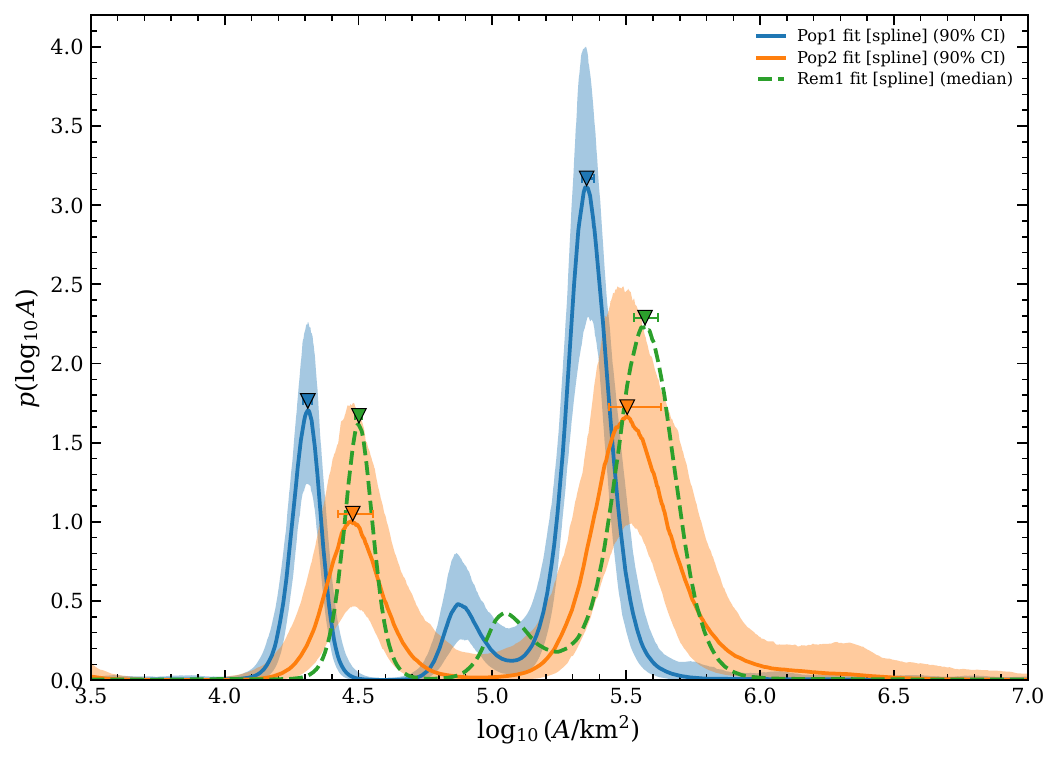}
\includegraphics[width=0.48\textwidth]{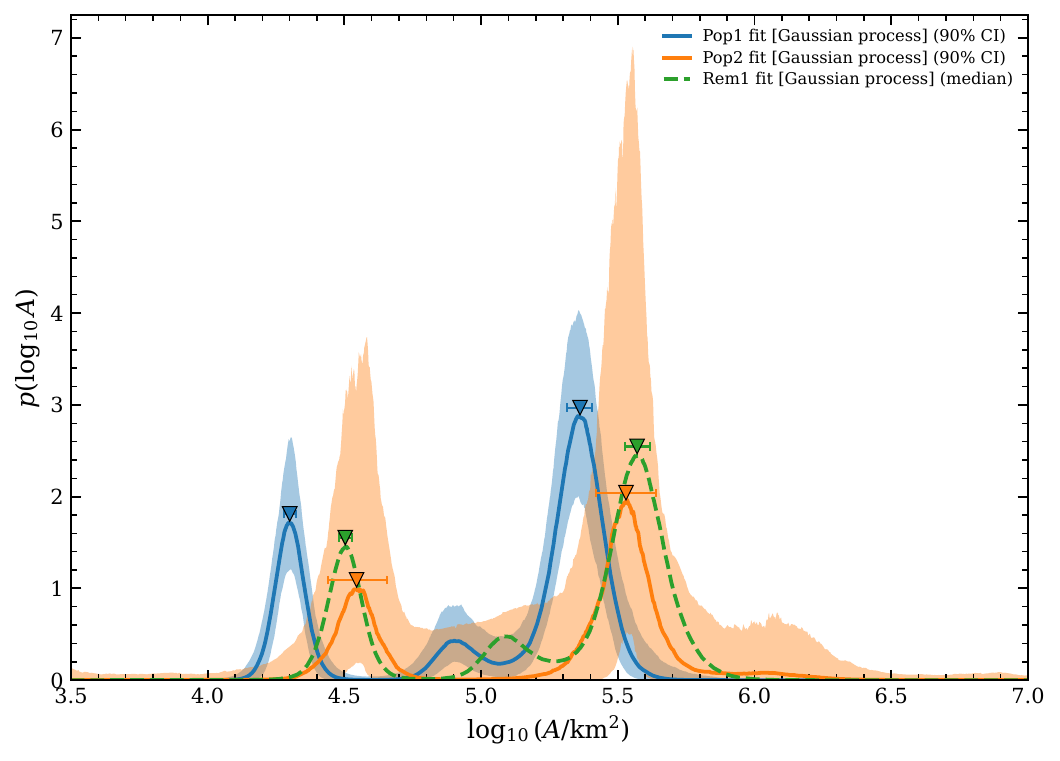}
\caption{\textbf{Non-parametric population distributions of the horizon-area statistic.}
Densities of $x=\log_{10}(A/{\rm km^2})$ for Pop1 (blue; pre-merger total areas), Pop2 (orange; areas of 2g BHs), and Rem1 (green; GR-predicted Pop1 remnant areas), reconstructed with two independent non-parametric schemes instead of a parametric mixture.
Solid curves are the posterior medians, bands the $90\%$ credible intervals, and triangles with horizontal error bars mark the corresponding quantities for the two peak locations.
\emph{Left}: spline model, with 20 nodes for Pop1, 10 nodes for Pop2, and 20 nodes for Rem1; \emph{right}: Gaussian-process model, with 60 nodes for each of Pop1, Pop2, and Rem1.}
\label{fig:nonp-ppc}
\end{figure}

\begin{figure}
\centering
\includegraphics[width=0.48\textwidth]{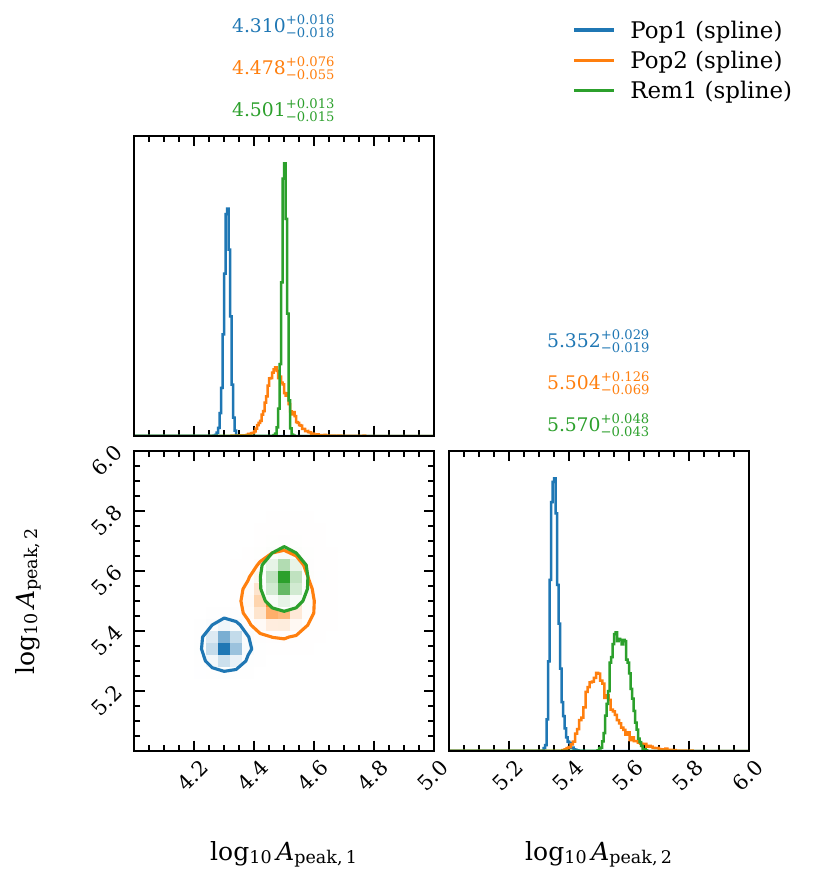}
\includegraphics[width=0.48\textwidth]{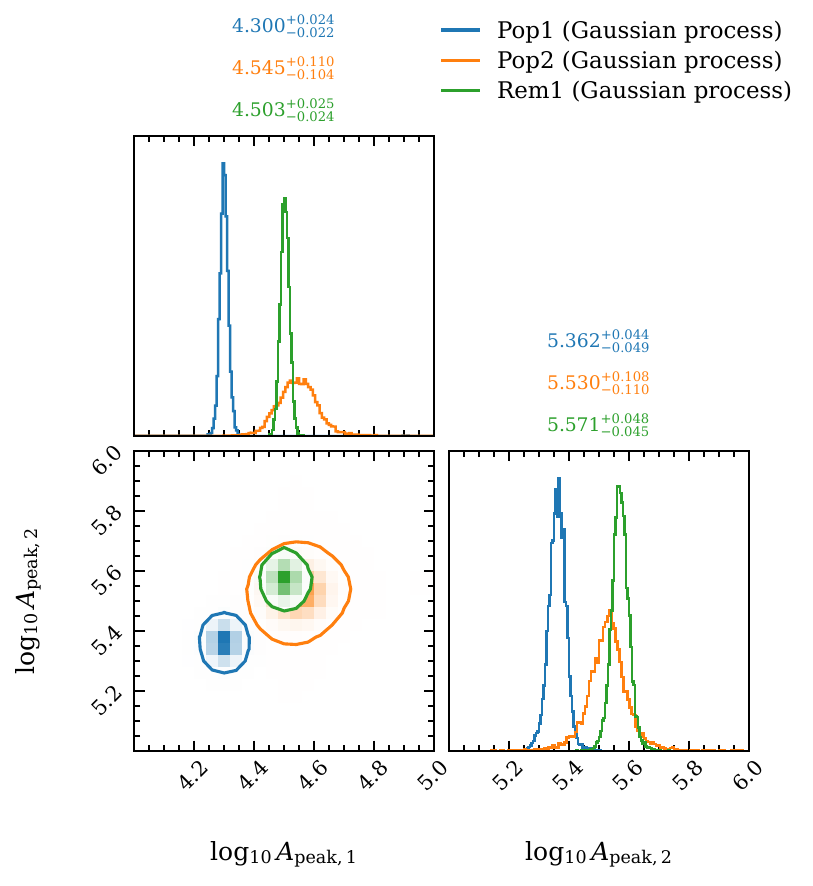}
\caption{\textbf{Joint posteriors of the peak locations from the non-parametric fits.}
Joint posteriors of $(\log_{10}A_{\rm peak,1},\log_{10}A_{\rm peak,2})$ of the Kerr horizon-area distributions for Pop1 (blue), Pop2 (orange), and the GR-predicted Pop1 remnants Rem1 (green), as in Figure~\ref{fig:para-peak} but for the two non-parametric schemes.
Contours enclose $90\%$ credible regions; titles give medians with $90\%$ credible intervals.
\emph{Left}: spline model, with 20 nodes for Pop1, 10 nodes for Pop2, and 20 nodes for Rem1; \emph{right}: Gaussian-process model, with 60 nodes for each of Pop1, Pop2, and Rem1.}
\label{fig:nonp-peak}
\end{figure}

\begin{figure}
\centering
\includegraphics[width=0.96\textwidth]{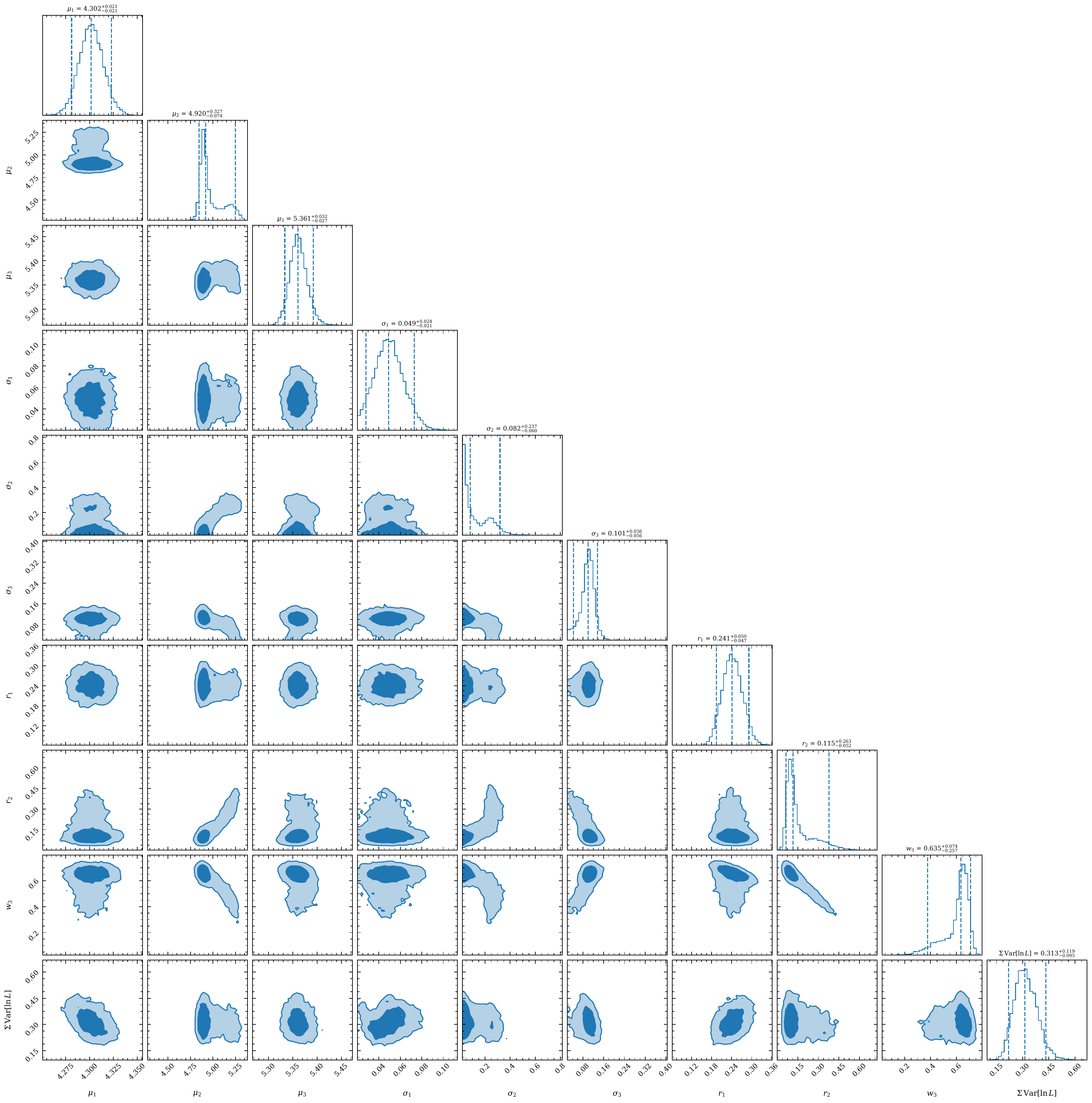}
\caption{\textbf{Hyperposterior of the Pop1 mixture fit.}
Three-component Gaussian-mixture fit to the Pop1 pre-merger total-horizon-area
statistic $x=\log_{10}(A/{\rm km^2})$. Contours enclose the $50\%$ and $90\%$
credible regions; titles give medians with $90\%$ credible intervals. The rightmost
column shows the total Monte Carlo variance of the hierarchical log-likelihood
estimate, which stays well below the penalty threshold throughout.}
\label{fig:pop1-corner}
\end{figure}

\begin{figure}
\centering
\includegraphics[width=0.96\textwidth]{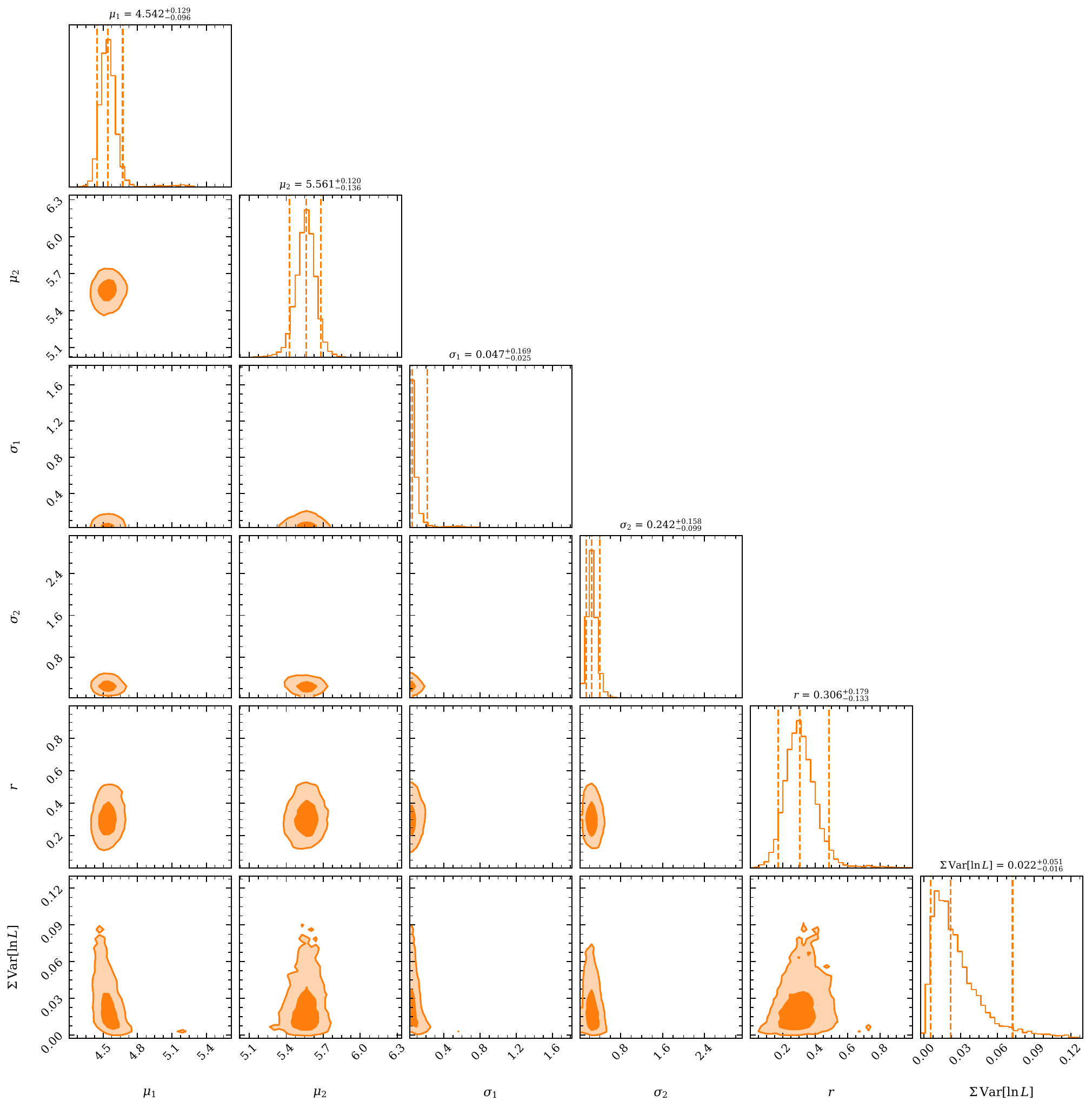}
\caption{\textbf{Hyperposterior of the Pop2 mixture fit.}
Same as Extended Data Figure~\ref{fig:pop1-corner}, but for the two-component
Gaussian-mixture fit to the Pop2 primary-horizon-area statistic.}
\label{fig:pop2-corner}
\end{figure}

\begin{figure}
\centering
\includegraphics[width=0.96\textwidth]{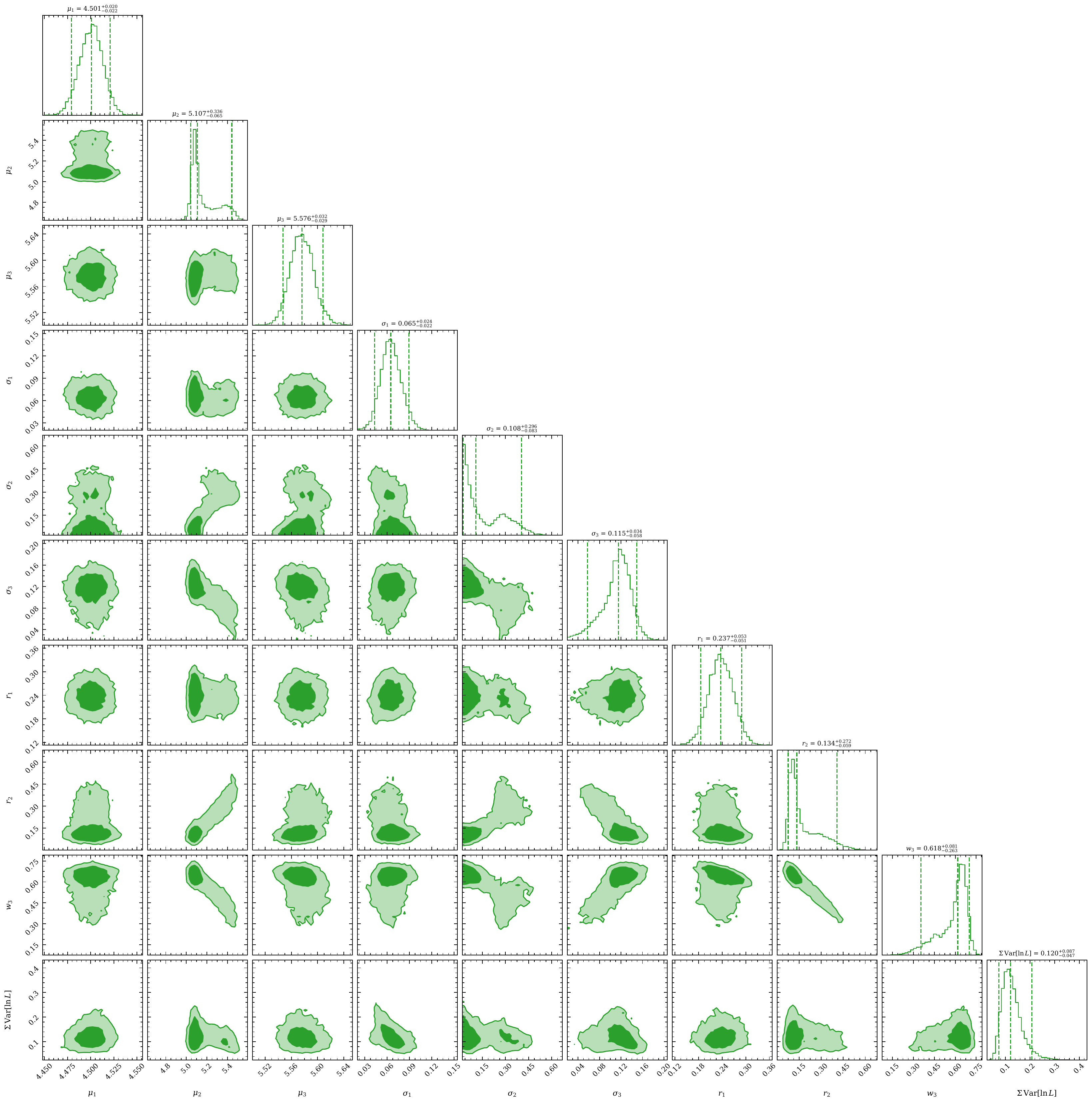}
\caption{\textbf{Hyperposterior of the Rem1 mixture fit.}
Same as Extended Data Figure~\ref{fig:pop1-corner}, but for the three-component
Gaussian-mixture fit to the Pop1 predicted remnant horizon-area statistic.}
\label{fig:rem1-corner}
\end{figure}

\end{document}